\documentstyle[twoside,graphicx]{elsart}
\graphicspath{{Figures/}{.}}
\begin{document} 
\begin{frontmatter}
  \title{\bf Kinematic Equations for Front Motion and Spiral-Wave
    Nucleation} \author[CNLS]{Aric Hagberg\thanksref{ARIC}}
  \author[BOKER]{Ehud Meron\thanksref{EHUD}}
  \address[CNLS]{Center for Nonlinear Studies and T-7, Theoretical Division,\\
    Los Alamos National Laboratory, Los Alamos, NM 87545}
  \thanks[ARIC]{\tt aric@lanl.gov}
  
  \address[BOKER]{The Jacob Blaustein
    Institute for Desert Research and the Physics Department,\\
    Ben-Gurion University, Sede Boker Campus 84990, Israel}
  \thanks[EHUD]{\tt ehud@bgumail.bgu.ac.il}
  
  \date{\today}

\begin{abstract}
  
  We present a new set of kinematic equations for front motion in
  bistable media. The equations extend earlier kinematic approaches by
  coupling the front curvature with the order parameter associated
  with a parity breaking front bifurcation.  In addition to naturally
  describing the core region of rotating spiral waves the equations
  can be be used to study the nucleation of spiral-wave pairs along
  uniformly propagating fronts.  The analysis of spiral-wave nucleation
  reduces to the simpler problem of droplet, or domain, nucleation in
  one space dimension.
\end{abstract}
\end{frontmatter}

%%%%%%%%%%%%%%%%%%%%%%%%%%%%%%%%%%%%%%%%%%%%%%%%%%%%%%%%%%%%%%%
%
%  Introduction
%
%%%%%%%%%%%%%%%%%%%%%%%%%%%%%%%%%%%%%%%%%%%%%%%%%%%%%%%%%%%%%%%
\vspace{-0.5in}
\section{Introduction} 
\vspace{-0.25in}

The onset of spatio-temporal disorder in reaction-diffusion systems is
often accompanied by the spontaneous nucleation of spiral waves.  The
conditions for spiral-wave nucleation, mechanisms behind it, and
implications on the resulting dynamics have been studied
extensively~\cite{CoWi:91,HoPa:91,Karma:93,Be:93,BFHNEE:94,HaMe:94b}.
Yet detailed studies of the nucleation process itself are still
lacking.  One difficulty in carrying out such studies stems from the
two-dimensional structures of spiral waves.  In this paper we report
on the derivation of new kinematic equations for front motion in
bistable media which reduce the two-dimensional spiral-wave nucleation
problem to a one-dimensional droplet nucleation problem~\cite{GuDr:83,Fife:79}.
We use the kinematic equations to demonstrate the destabilization of traveling
V-shape solutions to spiral nucleation.

\begin{figure}
  {\center \includegraphics[width=2.7in]{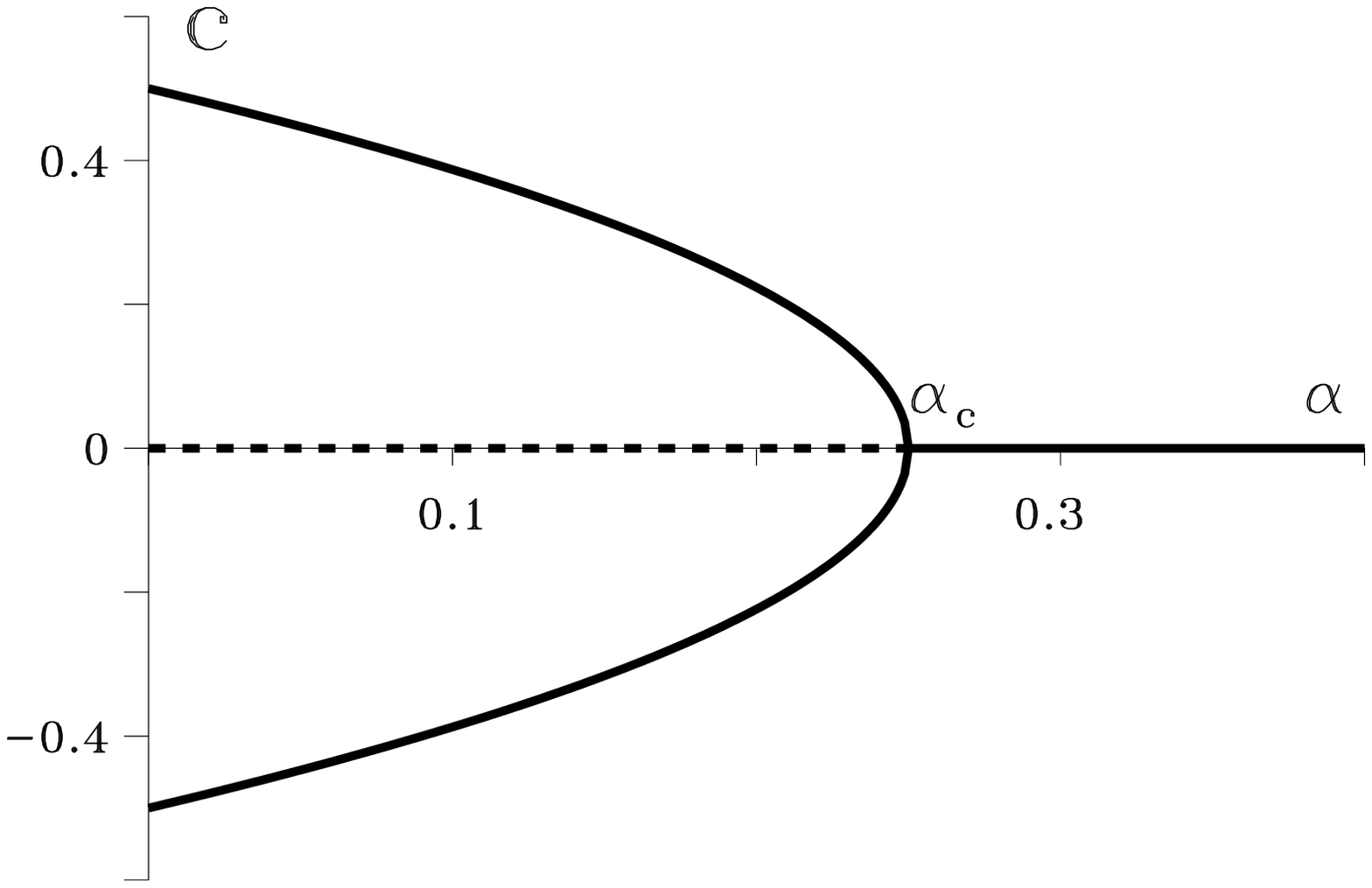}
    \includegraphics[width=2.7in]{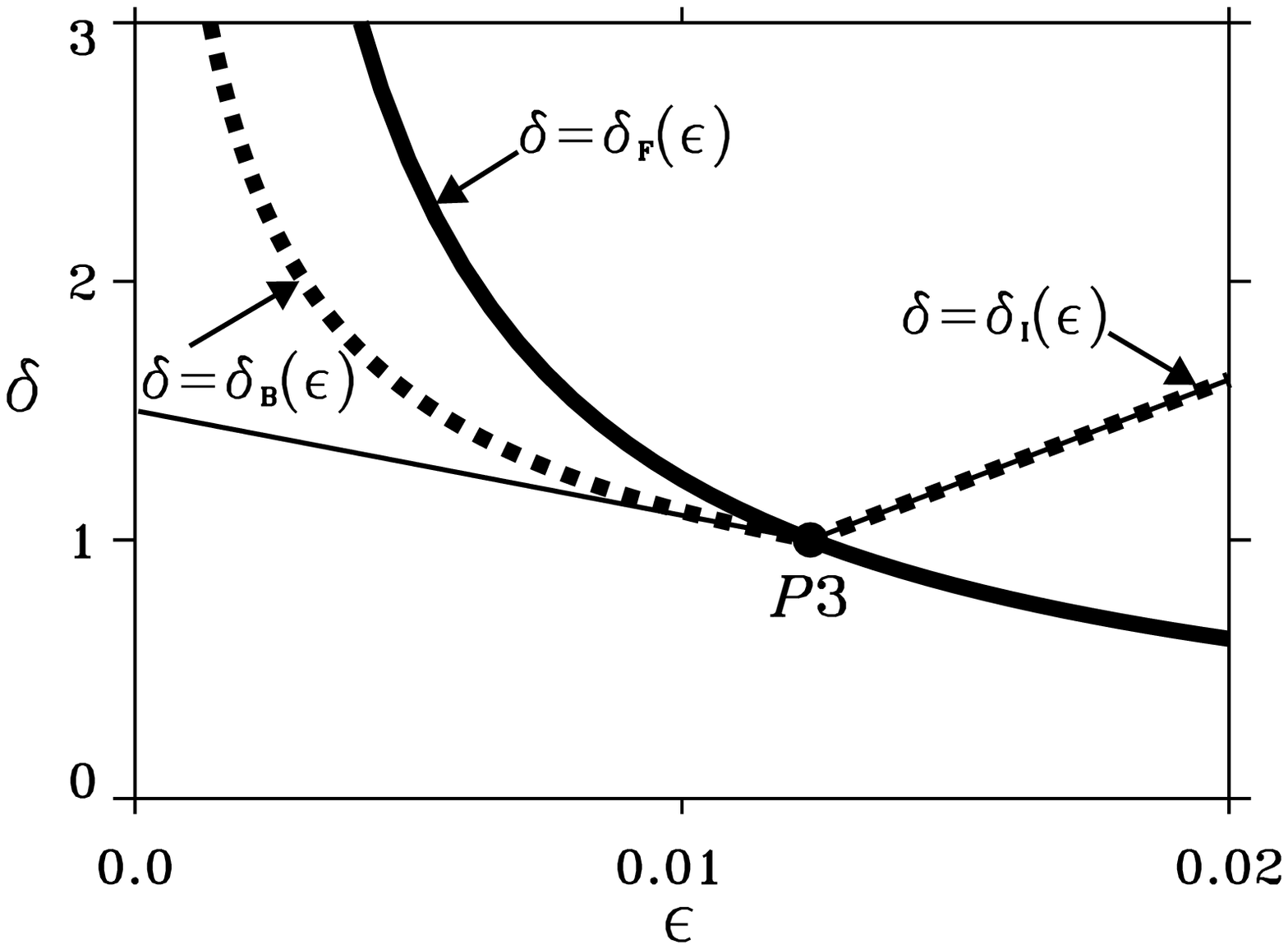} }
  \caption{
    (a) The nonequilibrium Ising-Bloch (NIB) front bifurcation.  The
    solid line represents a branch of stable front solutions with
    speed $c$.  At the bifurcation point, $\alpha=\alpha_c$, the
    stationary solution becomes unstable to a pair of
    counterpropagating fronts (b) The NIB front bifurcation and planar
    front transverse instability boundaries in the $\epsilon-\delta$
    parameter plane.  The thick line is the NIB bifurcation,
    $\delta_F(\epsilon)$, and the dashed lines are the boundaries for
    the transverse instability of Ising, $\delta_I(\epsilon)$, and
    Bloch, $\delta_B(\epsilon)$, fronts.  Above these lines planar
    fronts are unstable to transverse perturbations.  The thin lines
    are the approximations to the transverse instability boundaries
    obtained from the kinematic equations.  Parameters: $a_1=4.0$,
    $a_0=0$.}
  \label{fig:NIB}
\end{figure}

The derivation is carried out in a parameter range that includes a
pitchfork front bifurcation called a nonequilibrium
Ising-Bloch (NIB) bifurcation~\cite{CLHL:90,IMN:89,HaMe:94a,BRSP:94}.
At a NIB bifurcation, a stationary ``Ising'' front loses stability to
a pair of counter-propagating ``Bloch'' fronts as a parameter
$\alpha$ is decreased past a critical value (Fig.~\ref{fig:NIB}a).
The coexistence of the two Bloch fronts beyond the bifurcation
allows for nucleation events where a front line segment of 
one Bloch type undergoes a transition to the other Bloch
front.

The kinematic description of front motion consists of the following
equations:
\begin{itemize}
\item An equation for the order parameter, $C_0$, associated with the
  NIB bifurcation:
\begin{equation}
  {\partial C_0\over\partial t}=(\alpha_c-\alpha)C_0 - \beta C_0^3
  +\gamma\kappa + \gamma_0 +{\partial^2 C_0\over \partial s^2} -
  {\partial C_0 \over \partial s} \int_0^s \kappa C_n ds^\prime\,.
\label{C0}
\end{equation}
\item A geometric equation for the front curvature, $\kappa$:
\begin{equation}
  {\partial\kappa\over\partial t} = -(\kappa^2 +
  {\partial^2\over\partial s^2})C_n - {\partial\kappa\over\partial
    s}\int_0^s \kappa C_n ds^\prime \,.
\label{K}
\end{equation}
\item An equation relating the normal front velocity $C_n$, the
  curvature $\kappa$, and the order parameter $C_0$:
\begin{equation}
  C_n = C_0 - D\kappa\,.
\label{Cn}
\end{equation}
\end{itemize}
In these equations $s$ is the front arc length and the critical
parameter value $\alpha_c$ designates the NIB bifurcation point. The
kinematic equations are derived for a FitzHugh-Nagumo model with a
diffusing inhibitor. They generalize an earlier kinematic
approach~\cite{Mikhailov:90,Meron:92} by treating $C_0$ as an
independent dynamic mode rather than a constant.

The order parameter equation (\ref{C0}) yields the bifurcation diagram
of Fig.~\ref{fig:NIB}a for planar (not curved) front
solutions in a symmetric system ($\gamma_0=0$). The two Bloch branches
pertain to high and low $C_0$ values,
$C_0=C_0^\pm\equiv\pm\sqrt{(\alpha_c-\alpha)/\beta}$.
Because of the coupling between the order parameter and
curvature equations, curvature variations are capable of nucleating
low $C_0$ segments (droplets) in regions of high $C_0$. 
Droplet nucleation in the
kinematic equations corresponds to spiral-wave pair nucleation in the
physical two-dimensional plane.

Spatially uniform front transitions induced by curvature, producing for
example breathing spots, have been studied in
Ref.~\cite{HMRZ:97,HaMe:96c}.  The present paper is an extension of
this earlier work to nonuniformly curved fronts.  Complementary
results to those reported here appear in Ref.~\cite{HaMe:97}.

%%%%%%%%%%%%%%%%%%%%%%%%%%%%%%%%%%%%%%%%%%%%%%%%%%%%%%%%%%%%%%%
%
%  Derivation of the kinematic eqs
%
%%%%%%%%%%%%%%%%%%%%%%%%%%%%%%%%%%%%%%%%%%%%%%%%%%%%%%%%%%%%%%%
\section{Derivation of the kinematic equations} 
\vspace{-0.25in} The curvature equation~(\ref{K}) follows from purely
geometric
considerations~\cite{Mikhailov:90,Meron:92,Zykov:87,MePe:88,Brazhnik:96}.
The normal velocity relation~(\ref{Cn}) and the order parameter
equation~(\ref{C0}) can be derived using the FitzHugh-Nagumo
reaction-diffusion model with a diffusing inhibitor,
\begin{eqnarray}
  {\partial u\over\partial t}&=&
  \epsilon^{-1}(u-u^3-v)+\delta^{-1}\nabla^2u\,,\nonumber \\
  {\partial v\over\partial t}&=& u-a_1v-a_0+\nabla^2 v\,, \label{FHN}
\end{eqnarray} 
where $u$ and $v$, the activator and the inhibitor, are real scalar
fields and $\nabla^2$ is the Laplacian operator in two dimensions. The
parameter $a_1$ is chosen so that~(\ref{FHN}) describes a medium with
two stable spatially uniform states: an up state $(u_+,v_+)$ and a
down state $(u_-,v_-)$.  Ising and Bloch front solutions connect the
two uniform states $(u_\pm,v_\pm)$ as the spatial coordinate normal to
the front goes from $-\infty$ to $+\infty$.  The remaining parameter
space is spanned by $\epsilon, \delta$ and $a_0$, or alternatively by
$\eta=\sqrt{\epsilon\delta}$, $\mu=\epsilon/\delta$, and $a_0$.  Note
the parity symmetry $(u,v)\to(-u,-v)$ of~(\ref{FHN}) for $a_0=0$.

The NIB bifurcation line in the $\epsilon-\delta$ plane for $a_0=0$ is
shown in Fig.~\ref{fig:NIB}b.  For $\mu\ll 1$ it is given by
$\delta=\delta_F(\epsilon)=\eta_c^2/\epsilon$, or $\eta=\eta_c$, where
$\eta_c=\frac{3}{2\sqrt{2}q^3}$ and $q^2=a_1+1/2$~\cite{HaMe:94a}.
The single stationary Ising front that exists for $\eta>\eta_c$ loses
stability to a pair of counter-propagating Bloch fronts at
$\eta=\eta_c$.  Beyond the bifurcation ($\eta<\eta_c$) a Bloch front
of an up state invading a down state coexists with another Bloch front
of a down state invading an up state. Also shown in
Fig.~\ref{fig:NIB}b are the transverse instability boundaries (for
$a_0=0$), $\delta=\delta_I(\epsilon)= \epsilon/\eta_c^2$ and
$\delta=\delta_B(\epsilon)=\eta_c/\sqrt\epsilon$, for Ising and Bloch
fronts respectively.  Above these lines, $\delta>\delta_{I,B}$, planar
fronts are unstable to transverse
perturbations~\cite{HaMe:94b,HaMe:94c}.  All three lines meet at a
codimension 3 point $P3$: $\epsilon=\eta_c^2$, $\delta=1$, $a_0=0$.

The following assumptions are used in deriving Eqs.~(\ref{Cn})
and~(\ref{C0}): $\mu=\epsilon/\delta\ll 1$, the front speed $c$ is
small, the front is weakly curved, and curvature variations along the
front are weak. The second assumption is met by considering nearly
symmetric ($|a_0|<<1$) systems and restricting parameter values to the
Ising regime or the vicinity of the NIB bifurcation where the front
speed is $c\propto\sqrt{\eta_c-\eta}\ll 1$. The last two assumptions
are met by considering parameter values below or just beyond the
threshold for the transverse instability of fronts.

The kinematic equations are derived by transforming to an orthogonal 
coordinate system that moves with the
front and using singular perturbation theory with 
$\mu=\epsilon/\delta\ll 1$ as a small parameter~\cite{HMRZ:97}.  
The analysis of the narrow
front core region, where $u$ changes on a scale of order $\sqrt{\mu}$,
gives equation (\ref{Cn}) with
\begin{equation}
  C_0=-{3\over\eta\sqrt 2}v_f(s,t)\,, \qquad D=\delta^{-1}\,,
\label{rt}
\end{equation}
where $v_f(s,t)$ is the approximately constant value (in the direction normal
to the front) of the inhibitor $v$ in this region.

Away from the front core, where both $u$ and $v$ vary on a scale of
order unity, a free boundary problem is obtained by using the
solutions $u_\pm(v)$ of $u-u^3-v=0$ in the inhibitor ($v$) equation.
Matching the solutions from each side of the front at the front core
(in the limit $\mu\to 0$) gives the equation
\begin{eqnarray}
  {\partial v_f\over\partial t}& =& {\sqrt 2 (\eta_c-\eta)\over
    q\eta_c^2}v_f
  -{3\over 4\eta_c^2}v_f^3 - {4\over 3}a_0 \nonumber \\
  &&\mbox{}-{2(1-\delta^{-1})\over 3q}\kappa + {\partial^2 v_f\over
    \partial s^2}-{\partial v_f\over\partial s} \int_0^s \kappa C_n
  ds^\prime \ .
\label{vfeqn}
\end{eqnarray}

Equation~(\ref{vfeqn}) coincides with~(\ref{C0}) using the following
identifications: $C_0=-{3\over\eta\sqrt 2}v_f$, $\alpha={\eta\sqrt
  2\over q\eta_c^2}$, $\alpha_c={\sqrt 2\over q\eta_c}$, $\beta=1/6$,
$\gamma=\alpha_c(1-\delta^{-1})$, and $\gamma_0=2\alpha_cqa_0$.
More details about the derivation appear in Ref.~\cite{HaMe:97}.

%%%%%%%%%%%%%%%%%%%%%%%%%%%%%%%%%%%%%%%%%%%%%%%%%%%%%%%%%%%%%%%
%
%  Numerical solutions of the kinematic equations
%
%%%%%%%%%%%%%%%%%%%%%%%%%%%%%%%%%%%%%%%%%%%%%%%%%%%%%%%%%%%%%%%
\section{Numerical solutions of the kinematic equations}
\vspace{-0.25in}
\subsection{Spiral waves}
\begin{figure}
  \centering\includegraphics[width=4.0in]{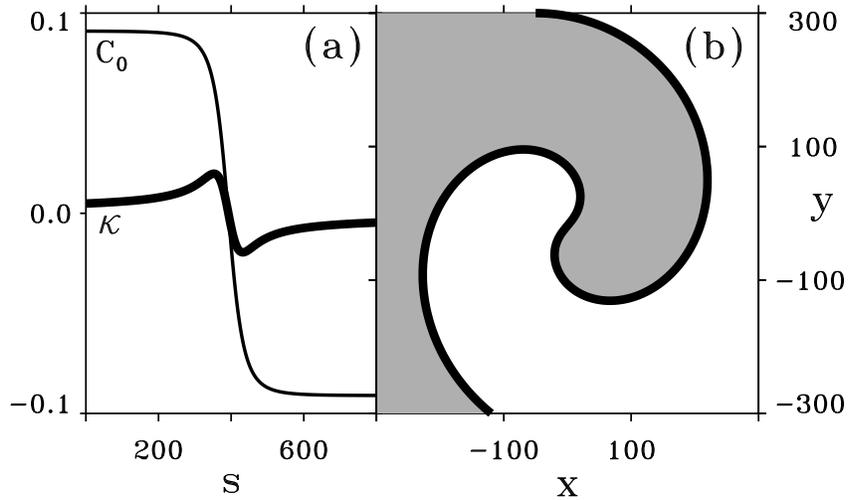}
  \caption{
    A front solution to the kinematic equations
    (\protect\ref{C0})-(\protect\ref{Cn}).  (a) The order parameter
    $C_0$ and the curvature $\kappa$ along the arc length $s$.  (b) In
    the $x-y$ plane the front solution corresponds to a rotating
    spiral wave. The shaded (light) region corresponds to an up (down)
    state.  Parameters: $a_1=4.0$, $a_0=0$, $\epsilon=0.01234$,
    $\delta=1.0.$ }
  \label{fig:spirals}
\end{figure}

\vspace{-0.25in} Consider a front solution in the kinematic equations
that connects the state $C_0=C_0^+$, $\kappa=0$, at $s=-\infty$ with
the state $C_0=C_0^-$, $\kappa=0$, at $s=+\infty$, where for a
symmetric model ($a_0=0$ or $\gamma_0=0$)
$C_0^\pm=\pm\sqrt{(\alpha_c-\alpha)/\beta}$.  Fig.~\ref{fig:spirals}a
shows such a solution obtained by numerically
integrating~(\ref{C0})-(\ref{Cn}).  As demonstrated in
Fig.~\ref{fig:spirals}b this front solution of the kinematic
equations~(\ref{C0})-(\ref{Cn}) represents a {\em spiral-wave} in the
FitzHugh-Nagumo model~(\ref{FHN}).  Unlike earlier kinematic
approaches~\cite{Mikhailov:90,Meron:92} the spiral core is naturally
captured.

\subsection{Traveling V-shape solutions and spiral-wave nucleation}
\vspace{-0.25in} For $C_0$ {\em constant}, the curvature
equation~(\ref{K}) and the normal velocity relation~(\ref{Cn}) have a
family of traveling V-shape solutions~\cite{Brazhnik:96,BrTy:96}
\begin{equation}
  \kappa(s)=-\frac{b^2}{C_0+C_n(0)\cosh(bs)}\,, \qquad
  b^2=C_n(0)^2-C_0^2 \,,
\label{V}
\end{equation}
where $C_n(0)=C_0-D\kappa(0)$ is an arbitrary constant. The traveling
V-shape solution~(\ref{V}) is also an exact solution of the kinematic
equations~(\ref{C0})-(\ref{Cn}) for $\delta=1$.  In that case
$\gamma=\alpha_c(1-\delta^{-1})=0$ and the order parameter equation
(\ref{C0}) decouples from the curvature equation for constant $C_0$
solutions. The specific values $C_0$ can assume are determined as
solutions of the cubic equation $(\alpha_c-\alpha)C_0-\beta
C_0^3+\gamma_0=0$.

\begin{figure}
  \centering\includegraphics[width=4.0in]{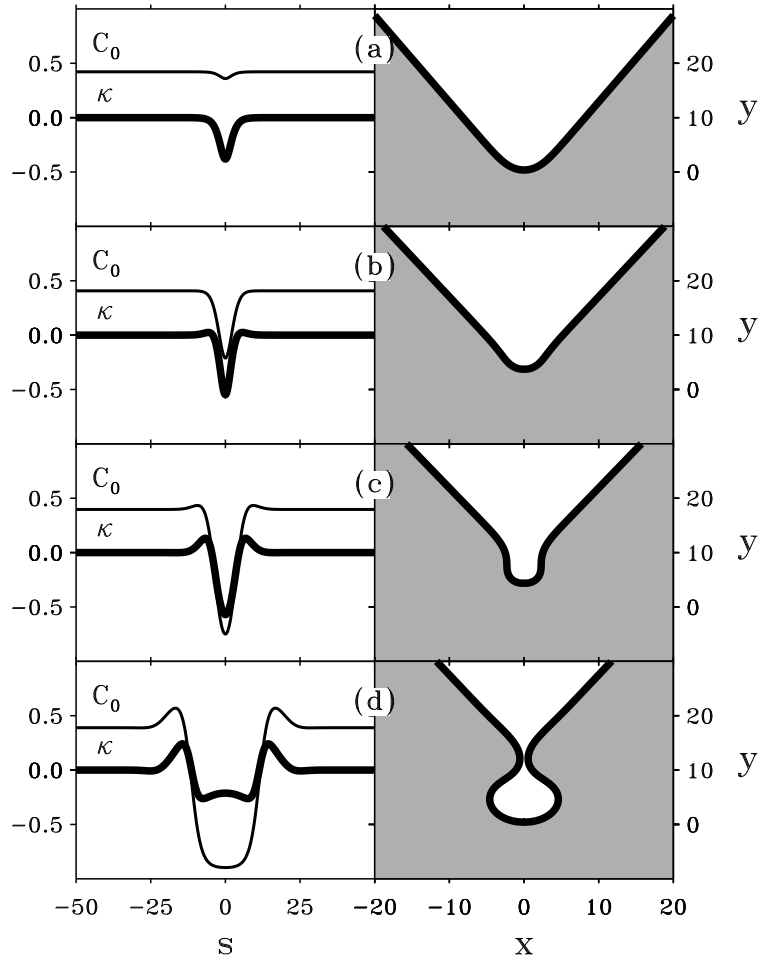}
  \caption{
    Nucleation of a spiral-wave pair from an initial traveling V-shape
    front.  Left column: the $C_0(s)$ and $\kappa(s)$ profiles.  Right
    column: the front line shape in the $x-y$ plane.  Parameters:
    $a_1=4.0$, $a_0=-0.0001$, $\epsilon=0.0115$, $\delta=1.063.$ }
  \label{fig:unstableV}
\end{figure}

We have also found stable traveling V-shape numerical solutions for
$\delta\ne 1$.  Fig.~\ref{fig:unstableV}a shows a V shape Bloch front
traveling stably at constant speed.  When approaching the NIB
bifurcation (by increasing $\epsilon$ and/or $\delta$) this solution
becomes unstable and nucleates spiral-wave pair.  The time evolution
of an unstable V-shape solution is shown in
Fig.~\ref{fig:unstableV}(b-d).

\begin{figure}
  \centering\includegraphics[width=5.5in]{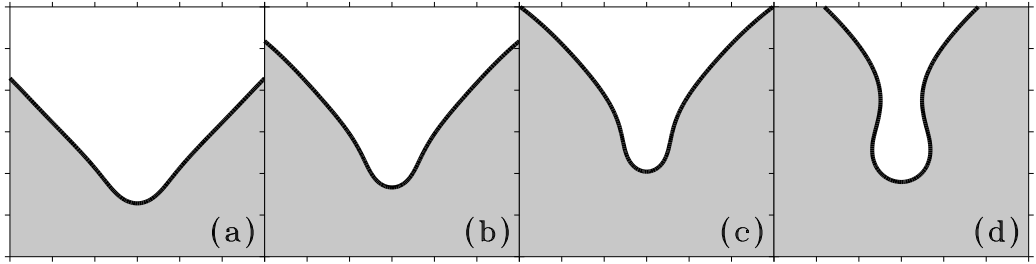}
  \caption{Spiral-wave nucleation from an unstable traveling
    V-shape solution of the FitzHugh-Nagumo
    equations~(\protect\ref{FHN}).  The initial traveling V-shape
    solution (a) slows down (b) at the high curvature region and
    nucleates a pair of spiral waves (c).  Depending on the parameters
    the resulting fronts may either repel or reconnect (d).
    Parameters: $\epsilon=0.0122$, $\delta=1.064$, $a_1=4.0$,
    $a_0=-0.0001$.  }
  \label{fig:FHNV}
\end{figure}

We carried out numerical solutions of the FitzHugh-Nagumo
model~(\ref{FHN}) to test the prediction, based on solutions of the
kinematic equations, that traveling V-shape solutions are destabilized
as the NIB bifurcation is approached.  Choosing parameter values close
to those of Fig.~\ref{fig:unstableV} an initial V form is indeed
unstable and a pair of spiral-waves nucleate in the region of highest
curvature as shown in Fig.~\ref{fig:FHNV}.  After the spirals form
the resulting traveling fronts may either repel each other or
reconnect and the V-shape solution will continue to propagate with the
possible nucleation of new spiral pairs.
Interactions between front segments are not included in the kinematic
equations~(\ref{C0})-(\ref{Cn}) so the resulting evolution of the
spiral pair and possible front reconnections cannot be captured.
Front interactions have been studied recently in the fast inhibitor
limit ($\eta\gg\eta_c$)~\cite{GMP:96}.
%%%%%%%%%%%%%%%%%%%%%%%%%%%%%%%%%%%%%%%%%%%%%%%%%%%%%%%%%%%%%%%
%
% Conclusion
%
%%%%%%%%%%%%%%%%%%%%%%%%%%%%%%%%%%%%%%%%%%%%%%%%%%%%%%%%%%%%%%%
\section{Conclusion}
\vspace{-0.25in} The kinematic equations presented in this paper
extend an earlier kinematic approach using the curvature
equation~(\ref{K}) and a linear relation between the normal velocity
and curvature~\cite{Mikhailov:90,Meron:92}.  Indeed, far into the
Bloch regime ($\alpha$ significantly smaller than $\alpha_c$), $C_0$
is no longer a slow degree of freedom and can be eliminated
adiabatically.  Equations~(\ref{C0}) and~(\ref{Cn}) then reduce to a
multivalued algebraic relation $C_0=C_0(\kappa)$ whose three branches
(two Bloch and one Ising) give approximately linear $C_n$ - $\kappa$
relations~\cite{HaMe:94c}.

But close to the NIB bifurcation $C_0$ becomes an active degree of
freedom and the earlier approach, often called the ``geometric
approach'', is no longer valid.  The new kinematic equations capture
the core structures of spiral waves and spiral-wave nucleation
processes.  Since one space dimension has been eliminated in deriving
the kinematic equations, the two-dimensional problem of spiral-wave
nucleation has been reduced to a simpler droplet nucleation problem in
one space dimension.

\section{Acknowledgements}
\vspace{-0.25in} We wish to thank Paul Fife for many interesting
discussions. This research was supported in part by grant No 95-00112
from the US-Israel Binational Science Foundation (BSF).

\end{document}